\documentclass{llncs}
\usepackage{graphicx}
\usepackage{amssymb}
\usepackage{amsmath}
\usepackage{booktabs}
\usepackage{hyperref}
\usepackage{bm}

\usepackage{todonotes}

\newcommand{\beq}{\begin{equation}}
\newcommand{\eeq}{\end{equation}}

\begin{document}

%%%%%%%%%%%%%%%%%%%%%%%%%%%%%%%%%%%%%%%%%%%%%%%%%%%%%%%%%%%%%%%%%%%%%%%%%%%%%%%%%%%%%%%%%%%%%%%%%%%%%%%%%%%%%%%%%%%%%%%%%%

% \title{Reconstruction of cortical surfaces, parcellation, and cortical thickness of clinical brain MRI scans with large spacing in the wild}
% \title{Cortical analysis of clinical brain MRI scans of any resolution and contrast}
\title{Cortical analysis of heterogeneous clinical brain MRI scans for large-scale neuroimaging studies}
% \title{ClinSurfer: Cortical surface analysis of \\ heterogeneous clinical MRI scans}
%\title{Cortical reconstruction, parcellation, and thickness of clinical brain MRI of any resolution and contrast}

%%%%%%%%%%%%%%%%%%%%%%%%%%%%%%%%%%%%%%%%%%%%%%%%%%%%%%%%%%%%%%%%%%%%%%%%%%%%%%%%%%%%%%%%%%%%%%%%%%%%%%%%%%%%%%%%%%%%%%%%%%

\author{Karthik Gopinath\thanks{Corresponding author: K. Gopinath. \textbf{Email:}~\email{kgopinath@mgh.harvard.edu}} \and
Douglas N. Greve\and Sudeshna Das \and Steve Arnold \and \\ Colin Magdamo \and Juan Eugenio Iglesias}
\institute{Athinoula A. Martinos Center for Biomedical Imaging, \\ Massachusetts General Hospital and Harvard Medical School}
% •	Karthik Gopinath
% •	Douglas Greve 
% •	Sudeshna Das 
% •	Steve Arnold 
% •	Colin Magdamo 
% •	Juan Eugenio Iglesias 

%%%%%%%%%%%%%%%%%%%%%%%%%%%%%%%%%%%%%%%%%%%%%%%%%%%%%%%%%%%%%%%%%%%%%%%%%%%%%%%%%%%%%%%%%%%%%%%%%%%%%%%%%%%%%%%%%%%%%%%%%%

\maketitle

\vspace{-3pt}

\begin{abstract}

Surface analysis of the cortex is ubiquitous in human neuroimaging with MRI, e.g., for cortical registration, parcellation, or thickness estimation. The convoluted cortical geometry requires isotropic scans (e.g., 1mm MPRAGEs) and good gray-white matter contrast for 3D reconstruction. This precludes the analysis of most brain MRI scans acquired for clinical purposes. Analyzing such scans would enable neuroimaging studies with sample sizes that cannot be achieved with current research datasets, particularly for underrepresented populations and rare diseases. Here we present the first method for cortical reconstruction, registration, parcellation, and thickness estimation for clinical brain MRI scans of any resolution and pulse sequence. The methods has a learning component and a classical optimization module. The former uses domain randomization to train a CNN that predicts an implicit representation of the white matter and pial surfaces (a signed distance function) at 1mm isotropic resolution, independently of the pulse sequence and resolution of the input. The latter uses geometry processing to place the surfaces while accurately satisfying topological and geometric constraints, thus enabling subsequent parcellation and thickness estimation with existing methods. We present results on 5mm axial FLAIR scans from ADNI and on a highly heterogeneous clinical dataset with 5,000 scans. Code and data are publicly available at \url{https://surfer.nmr.mgh.harvard.edu/fswiki/recon-all-clinical}.

% Leave this for the final paper, if accepted
% \keywords{Clinical MRI  \and cortical analysis \and thickness \and parcellation.}

\end{abstract}

\section{Introduction}
Clinical MRI exams account for the overwhelming majority of brain MRI scans acquired worldwide every year~\cite{oren2019curbing}. These exams comprise several scans acquired during a session with different orientations (axial, coronal, sagittal), resolutions, and MRI contrasts. The acquisition hardware and pulse sequence parameters differ significantly across (and even within) centers, leading to highly heterogeneous data. Since cortical thickness is a robust biomarker in the study of normal aging~\cite{salat2004thinning} and many brain disorders and diseases~\cite{querbes2009early,pereira2012assessment,rosas2002regional}, methods that can extract parcellations and thickness measurements from clinical scans (while registering to a reference spherical coordinate frame)  are highly desirable. However, cortical analysis of clinical scans is complex due to large slice spacing (resulting in incomplete cortex geometry description) and heterogeneous acquisitions (hindering supervised approaches leveraging image intensity distributions). 

% Existing neuroimaging research studies~\cite{hibar2018cortical} rely on isotropic scans with good gray-white matter contrast (typically a 1mm MPRAGE) and utilize prior information on  tissue intensities. Classical cortical analysis pipelines like FreeSurfer~\cite{dale1999cortical,fischl1999cortical} generate two triangle meshes per hemisphere, one for the white matter (WM) surface and one for the pial surface. This is often achieved by:  tessellating a WM mask generated from a volumetric segmentation algorithm; correcting the toplogy of the mesh to prevent holes and handles; further deforming this mesh to fit the WM surface while preventing self-intersections; and fitting the pial surface by letting the WM surface evolve outwards. The spherical topology of the surfaces enables mapping coordinates to a sphere enabling computation of vertex-wise statistics across subjects in a common coordinate frame. 

Existing neuroimaging research studies~\cite{hibar2018cortical} rely on isotropic scans with good gray-white matter contrast (typically a 1mm MPRAGE) and utilize prior information on tissue intensities. Classical cortical analysis pipelines like FreeSurfer~\cite{dale1999cortical,fischl1999cortical} generate two triangle meshes per hemisphere, one for the white matter (WM) surface and one for the pial surface, while preventing self-intersections. The spherical topology of the surfaces enables mapping coordinates to a sphere, thus  enabling computation of vertex-wise statistics across subjects in a common space. % coordinate frame. 

Over the last two years, machine learning approaches for cortical reconstruction on 1mm MPRAGEs have emerged. Methods based on signed distance functions (SDF) like DeepCSR~\cite{cruz2021deepcsr} or SegRecon~\cite{gopinath2021segrecon} predict voxel-wise SDFs for the WM and pial surfaces. The final meshes are computed as the SDF isosurfaces and do not guarantee topological correctness. PialNN~\cite{ma2021pialnn} uses an explicit representation to project the pial surface from the WM surface, which is assumed to be topologically correct. Approaches based on surface deformation like TopoFit~\cite{hoopes2021topofit} or Vox2Cortex~\cite{bongratz2022vox2cortex} use image and graph convolutions to predict a deformation that maps a topologically correct template mesh to an input MRI, thus generating WM and pial surfaces. However, these approaches neither prevent self intersections nor guarantee topological correctness. 

\noindent\textbf{Contribution:}
Our proposed method allows cortical analysis of brain MRI scans of any orientation, resolution, and MRI contrast without retraining, making it possible to use it out of the box for straightforward analysis of large datasets ``in the wild''. The proposed method combines two modules: a  convolutional neural network (CNN) that estimates SDFs of the WM and pial surfaces, and a classical geometry processing module that places the surfaces while satisfying geometric constraints (no self-intersections, spherical topology, regularity). The CNN capitalizes on recent advances in domain randomization to provide robustness against changes in acquisition -- in contrast with existing learning approaches that can only process images acquired with the same resolution and MRI contrast as the scans they were trained on. Finally, our method's classical geometry processing gives us geometric guarantees and grants instant access to an array of existing methods for cortical thickness estimation, registration,  and parcellation~\cite{fischl1999cortical}. %(e.g.,  FreeSurfer). 

\noindent\textbf{Further related work:} 
% The parameterization of surfaces as SDFs is commonplace in the computer graphics literature and has been successfully combined with deep neural networks in several domains~\cite{park2019deepsdf}, including cortical reconstruction~\cite{cruz2021deepcsr}. Our robustness to MRI contrast and resolution changes builds on recent ideas from the domain randomization literature~\cite{tobin2017domain}, which have been successfully applied to MRI analysis~\cite{billot2021synthseg,hoffmann2021synthmorph}. In short, these techniques train supervised CNNs with synthetic images generated from segmentations on the fly at every iteration. Crucially, the simulation parameters (orientation, contrast, resolution) are randomly sampled from uniform distributions at every mini-batch (most often leading to unrealistic appearance), such that the CNN learns not to rely on these features and becomes agnostic to them.
The parameterization of surfaces as SDFs has been combined with deep neural networks in several domains~\cite{park2019deepsdf}, including cortical reconstruction~\cite{cruz2021deepcsr}. Our robustness to MRI contrast and resolution changes is achieved using ideas from the domain randomization literature~\cite{tobin2017domain}, which involves training supervised CNNs with synthetic images generated from segmentations on the fly at every iteration. These techniques have been successfully applied to MRI analysis~\cite{billot2021synthseg,hoffmann2021synthmorph} and use random sampling of simulation parameters such as orientation, contrast, and resolution from uniform distributions at every mini-batch, which results in unrealistic appearance, making the CNN agnostic to these features. 

%%%%%%%%%%%%%%%%%%%%%%%%%%%%%%%%%%%%%%%%%%%%%%%%%%%%%%%%%%%%%%%%%%%%%%

\section{Methods}

\begin{figure}[t]
\centering
\includegraphics[width=0.97\textwidth]{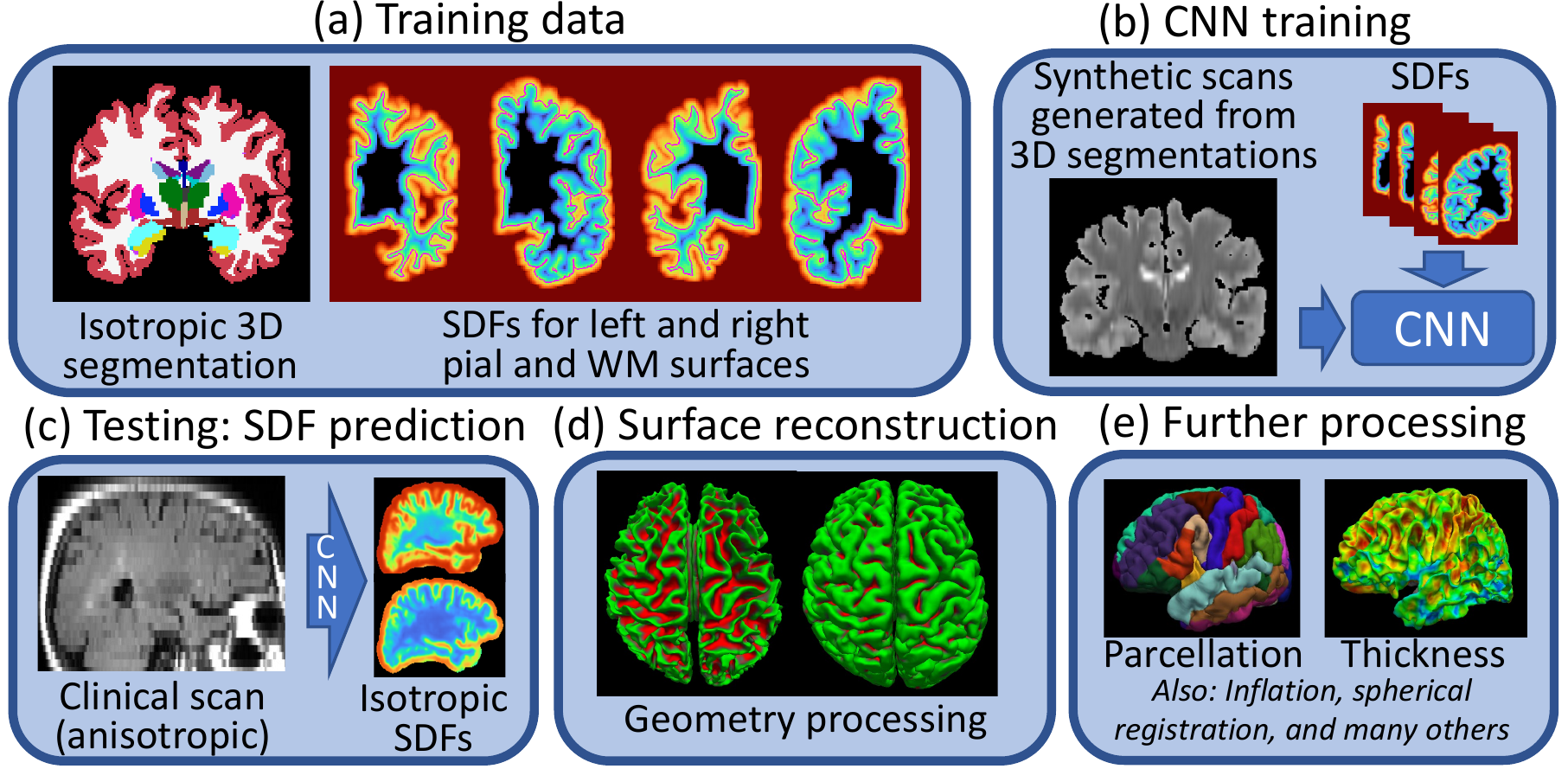}
%\vspace{-5pt}
\caption{
Overview of our proposed approach for cortical analysis of clinical brain MRI scans of any resolution and MRI contrast, without retraining. The images shown in (c-e) correspond to a real axial FLAIR scan with 5mm slice spacing and 5mm thickness.}\label{fig:overview} 
%\vspace{-5pt}
\end{figure}

Our proposed method (Figure~\ref{fig:overview}) has two distinct components: a learning module to estimate isotropic SDFs from anisotropic scans and a geometry processing module to place the WM and pial surfaces with topological constraints.

\subsection{Learning of SDFs}
\label{sec:learning}

% This module aims to estimate isotropic SDFs of WM and pial surfaces for both hemispheres in a contrast- and resolution-independent fashion. A domain randomization approach based on training a voxel-wise regression CNN with synthesizing a training dataset comprising volumetric segmentations and corresponding surfaces (real images are \emph{not} used) predicts SDFs. Such training data were obtained ``for free'' by running FreeSurfer on isotropic MPRAGEs from HCP~\cite{glasser2013minimal}.

This module estimates isotropic SDFs of the WM and pial surfaces of both hemispheres in a contrast- and resolution-independent fashion. It utilizes a domain randomization approach based on training a voxel-wise SDF regression CNN with synthetic data, which comprises volumetric segmentations and corresponding surfaces (real images are \emph{not} used). Such training data can be obtained ``for free'' by running FreeSurfer on isotropic T1 scans (we used the HCP dataset~\cite{glasser2013minimal}).

Given a 3D segmentation and four surface meshes (WM and pial surfaces for each hemisphere; see Figure~\ref{fig:overview}a), we compute the following input/target pairs at every training iteration. As input, we simulate a synthetic MRI scan of random orientation, resolution, and contrast from the 3D segmentation. For this purpose, we use a Gaussian mixture model conditioned on the (spatially augmented) labels, combined with models of bias field, resolution, and noise similar to~\cite{billot2021synthseg}. We use random sampling to determine the orientation (coronal, axial, sagittal, or isotropic), slice spacing (between 1 and 9mm), and slice thickness (between 1mm and the slice spacing). The thickness is simulated with a Gaussian kernel across slices. The final synthetic image is upscaled to 1mm isotropic resolution, such that the CNN  operates on input-output pairs of the same size and resolution.

As regression targets, we use voxel-wise SDFs computed from the WM and pial meshes for both hemispheres. The computation of the SDFs would greatly slow down CNN training if performed on the fly. Instead, we precompute them before training and deform them nonlinearly (along with the 3D segmentation) for geometric augmentation during training. While this is only an approximation to the real SDF, it respects the zero-level-set that implicitly defines the surface, and we found it to work well in practice. An example of a synthetic scan and target SDFs used to train the CNN are shown in Figure~\ref{fig:overview}b. 

The regression CNN is trained by feeding the synthetic images to the network and optimizing the weights to minimize the L1 norm of the difference between the ground truth and predicted distance maps. In practice, we clip the SDFs at an absolute value of 5mm to prevent the CNN from wasting capacity trying to model relatively small variations far away from the surfaces of interest. At test time, the input scan is upscaled to the isotropic resolution of the training data and pushed through the CNN to obtain the predicted SDFs.

\subsection{Geometry processing for surface placement}

To process real clinical scans, we first feed them to the trained CNN to predict the SDFs for the pial and WM surfaces for both hemispheres (Figure~\ref{fig:overview}c). 
To avoid generating topologically incorrect surfaces from these SDFs, we capitalize on the extensive literature on the geometry processing of cortical meshes with classical techniques. For reconstructing WM surfaces, we run SynthSeg~\cite{billot2021synthseg} on the input scan to obtain two binary masks corresponding to the left and right WM labels. From this point on, processing happens independently for each hemisphere. First, we fill in the holes in the hemisphere's mask and tessellate it to obtain the initial WM mesh. Then, we smooth the mesh and use automated manifold surgery~\cite{fischl2001automated} to guarantee spherical topology. Next, we iteratively deform the WM mesh by minimizing an objective function consisting of a fidelity term and a regularizer. 

Specifically: let $\mathcal{M} = (\bm{X}, \mathcal{K})$ denote a triangle mesh, where $\bm{X}=[\bm{x}_1, \ldots, \bm{x}_V]$ represents the coordinates of its $V$ vertices ($\bm{x}_v \in \mathbb{R}^3$), and $\mathcal{K}$ represents the connectivity. Let $D_w(\bm{r})$ be the SDF for the WM surface estimated by our CNN, where $\bm{r}$ is the spatial location. The objective function (``energy'') is the following:
\begin{align}
E[\bm{X}; D_w(\bm{r}), \mathcal{K}] = &  \sum_{v=1}^V [\tanh D_w(\bm{x}_v)]^2 
+ \lambda_1 \sum_{v=1}^V \sum_{u\in\mathcal{N}_v} [\bm{n}_v^t (\bm{x}_v - \bm{x}_u)]^2 \nonumber \\
+ & \lambda_2 \sum_{v=1}^V \sum_{u\in\mathcal{N}_v} \left\{  [\bm{e}_{1v}^t (\bm{x}_v - \bm{x}_u)]^2 + [\bm{e}_{2v}^t (\bm{x}_v - \bm{x}_u)]^2 \right\} . \label{eq:wm_placement} 
\end{align}
The first term in Equation~\ref{eq:wm_placement} is the fidelity term, which encourages the SDF to be zero on the mesh vertices; we squash the SDF through a $tanh$ function to prevent huge gradients far away from zero. The second and third terms are regularizers that endow the mesh with a spring-like behavior~\cite{dale1999cortical}: $\bm{n}_v$ is the surface normal at vertex $v$; $\bm{e}_{1v}$ and $\bm{e}_{2v}$ define an orthonormal basis for the tangent plane at vertex $v$; $\mathcal{N}_v$ is the neighborhood of $v$ according to $\mathcal{K}$; and $\lambda_1$ and $\lambda_2$ are relative weights, which we define according to~\cite{dale1999cortical} ($\lambda_1=0.0006$, $\lambda_2=0.0002$). Optimization is performed with gradient descent. At every iteration, self-intersections are monitored and eliminated by reducing the step size as needed~\cite{dale1999cortical}. 

The pial surface is fitted with a very similar procedure, but using the predicted SDF of the pial surface. Figure~\ref{fig:overview}d shows examples of reconstructed surfaces for the axial FLAIR scan from Figure~\ref{fig:overview}c. 
Given the fitted WM and pial surfaces, we use  FreeSurfer to compute cortical thickness, parcellation, and registration to a common coordinate frame in spherical coordinates (Figure~\ref{fig:overview}e).

\subsection{Implementation details}
% \noindent\textbf{Implementation details}
Our voxel-wise regression CNN is a 3D U-net~\cite{ronneberger2015u} trained with synthetic pairs generated on the fly as explained in Section~\ref{sec:learning} above. The U-net has 5 levels with 2 layers each, uses 3$\times$3$\times$3 convolutions and exponential linear activations. The layers have $24^l$ features, where $l$ is the level number. The last layer uses linear activation functions to model the SDFs. The CNN weights were optimized with stochastic gradient descent using a fixed step size of 0.0001 and 300,000 iterations (enough to converge). At test time, the run time is dominated by the geometry processing (2-3 hours, depending on the complexity of the manifold surgery).

%%%%%%%%%%%%%%%%%%%%%%%%%%%%%%%%%%%%%%%%%%%%%%%%%%%%%%%%%%%%%%%%%%%%%%

\section{Experiments and Results}

\subsection{Datasets}

\noindent- \textbf{HCP}: we used 150 randomly selected subjects (71 males, ages: 29.9$\pm$3.4 years) from HCP~\cite{glasser2013minimal} to train the U-net. We ran FreeSurfer to obtain the segmentations and SDFs (images are discarded as they are not used in training).

\noindent- \textbf{ADNI}: in our first experiment, we used 1mm MPRAGE and corresponding 5mm axial FLAIR scans of 200 randomly selected subjects from the ADNI dataset~\cite{jack2008alzheimer} (95 males, ages 74.5$\pm$7.4 years). This setup enables us to directly compare the results from research- and clinical-grade scans.

\noindent- \textbf{Clinical}: this dataset comprises 9,735 scans from 1,367 MRI sessions of distinct subjects with memory complaints (749 males, ages 18-90) from HOSPITAL. Surfaces were  successfully generated for 5,064 scans; the rest failed due to insufficient field of view. This dataset includes a wide range of MR contrasts and resolutions. We note that this dataset also includes 581 1mm MPRAGE scans.

The availability of 1mm MPRAGEs for some of the subjects enables us to process them with FreeSurfer and use the result as ground truth~\cite{iscan2015test} (Figure~\ref{fig:competing}a).

\subsection{Competing methods}

\begin{figure}[t]
\centering
\includegraphics[width=0.87\textwidth]{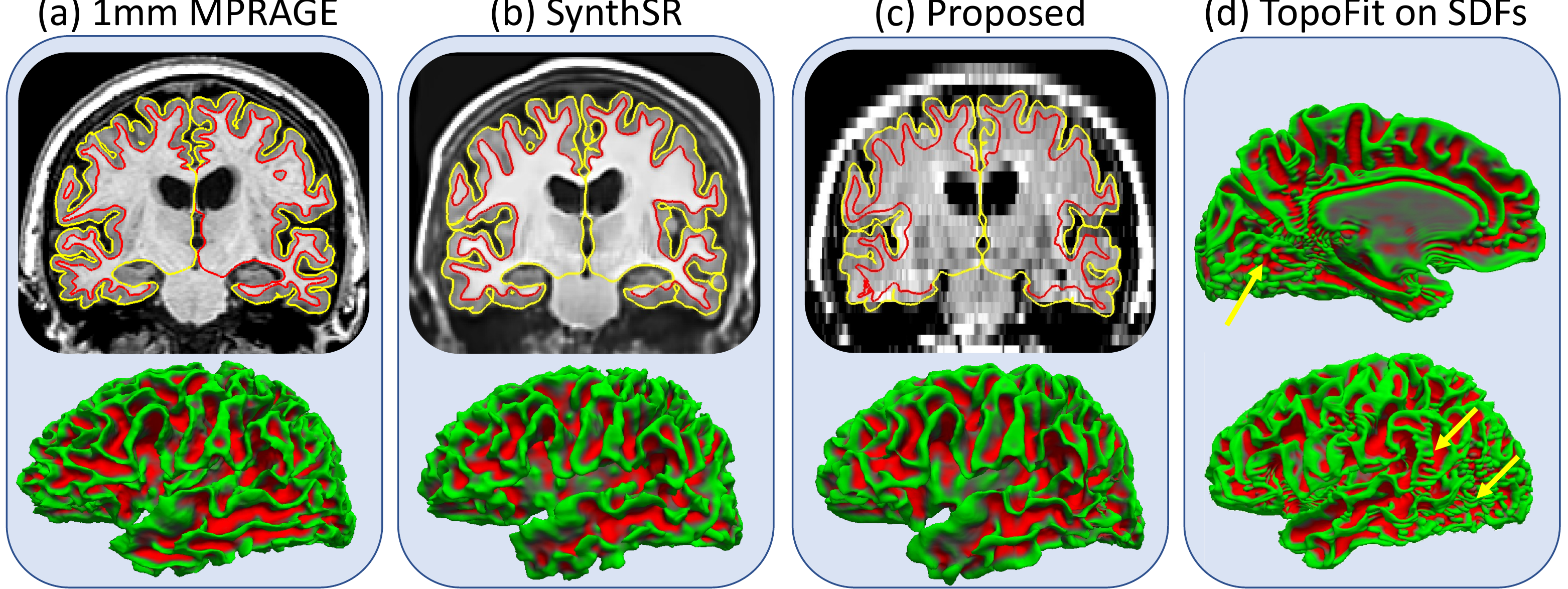}
%\vspace{-5pt}
%https://docs.google.com/drawings/d/1eEZYRcbn47vY7WjSreP7By_vtvRGjiirf-tZ3UWQmXM/edit?usp=sharing
% of competing methods 
\caption{
Qualitative comparison on a 5mm axial FLAIR scan from ADNI. (a)~Ground truth T1 with WM (red) and pial (yellow) surfaces estimated by FreeSurfer (top); and 3D reconstruction of the left WM surface (bottom); (b)~Synthetic T1 produced by SynthSR with FreeSurfer surfaces. (c)~Our proposed method. (d)~Examples of WM surfaces produced by TopoFit, trained to predict on the output of our SDF predictor, show small blobs in different areas (e.g., indicated by yellow arrows)}\label{fig:competing} 
%\vspace{-5pt}
\end{figure}

To the best of our knowledge, the only existing competing method for our proposed algorithm is SynthSR~\cite{iglesias2021joint}, which utilizes a synthetic data generator like ours to turn scans of any resolution and contrast into synthetic 1mm MPRAGES -- which can be subsequently processed with FreeSurfer to obtain surfaces (Figure~\ref{fig:competing}b). Compared with our proposed approach (Figure~\ref{fig:competing}c), this pipeline inherits the smoothness of the synthetic MPRAGE, leading to smoother surfaces that may miss larger folds. We also tried training TopoFit\cite{hoopes2021topofit} on the synthetic images and predicted SDFs, but failed to produce neural networks with good generalization ability, as they led to small blobs on the surfaces at test time (Figure~\ref{fig:competing}d).

\subsection{Results on the ADNI dataset}

\begin{figure}[th!]
\centering
\includegraphics[width=0.94\textwidth]{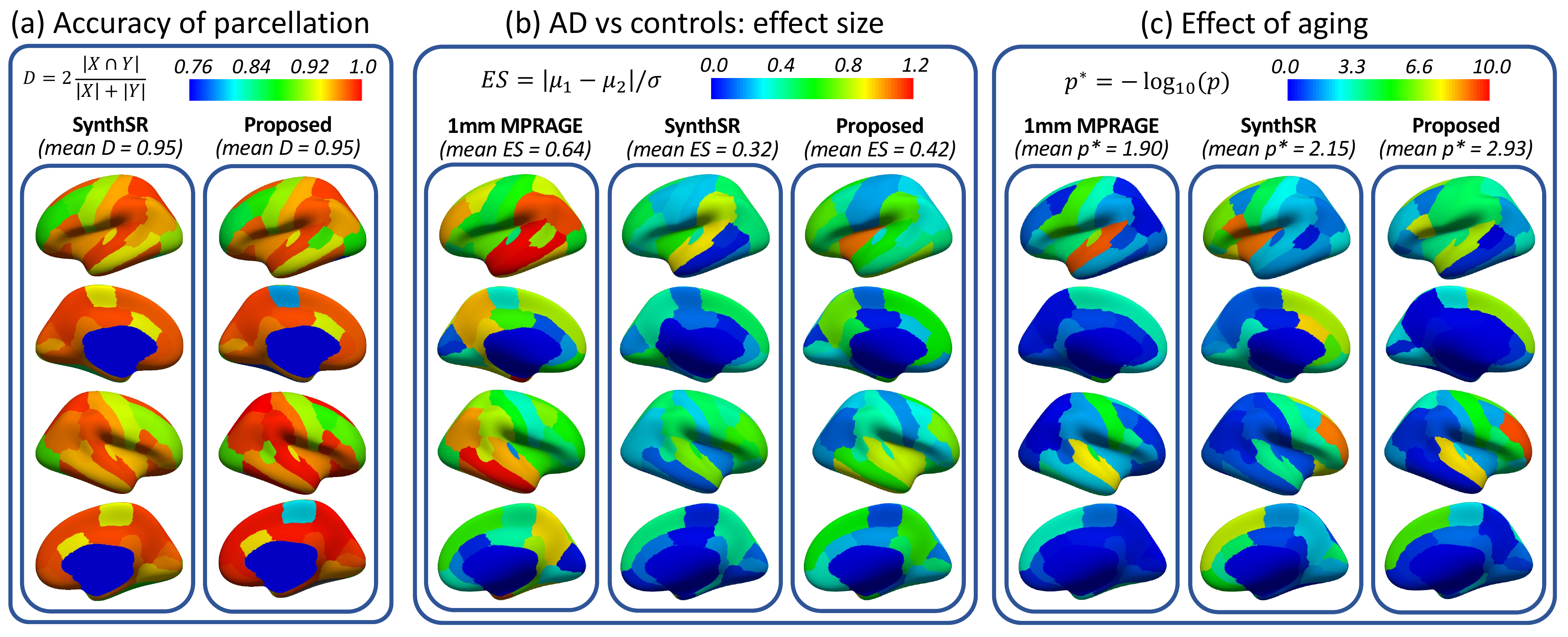}
%\vspace{-5pt}
%https://docs.google.com/drawings/d/1eEZYRcbn47vY7WjSreP7By_vtvRGjiirf-tZ3UWQmXM/edit?usp=sharing
\caption{
Summary of results on the ADNI dataset, displayed on the inflated surface of FreeSurfer's average subject (\textit{fsaverage}). (a)~Accuracy of parcellation for SynthSR and our proposed method using Dice scores. (b)~Ability to discriminate AD vs controls, measured with effect sizes. (c)~Effect of aging, measured as the strength of the (negative) correlation between age and thickness. The strength of the correlation is represented by p-values of a Student's $t$ test assessing whether the correlation significantly differs from zero; note that we log-transform the p-values for easier visualization.}\label{fig:ADNI} 
%\vspace{-5pt}
\end{figure}

Figure~\ref{fig:ADNI} summarizes the results on the ADNI dataset. 
%Rather than focusing on surface distance errors that are not easily interpretable (e.g., as in previous machine learning approaches), we evaluate our method using the performance on the downstream tasks that one is ultimately interested in. 
While previous machine learning approaches focus evaluation on distance errors, these can be difficult to interpret. Instead, we evaluate our method using the performance on the downstream tasks that one is ultimately interested in. 
First, we computed the accuracy of the Desikan-Killiany parcellation~\cite{desikan2006automated} produced by SynthSR and our proposed method. Figure~\ref{fig:ADNI}a shows the results on the inflated surface of the \textit{fsaverage} template. Since the parcellation is computed from the curvature of the WM surface, it is a relatively easy problem. The overlap between the ground truth parcellation and the two competing methods is very high. Dice scores over 0.90 are obtained for almost every region in both methods, and the average across regions is almost identical for both methods (0.95).

We then used the obtained parcellations to study the effect of Alzheimer's disease (AD) on cortical thickness, using a group study between AD subjects and elderly controls. For this purpose, we first fit a general linear model (GLM) to the cortical thickness at every parcel, using age, gender, and AD status as covariates. We then used the model coefficients to correct the thickness estimates for age and gender, and compared the thicknesses of the two groups. 

Figure~\ref{fig:ADNI}b shows the effect sizes (ES) for the reference 1mm MPRAGEs and the competing methods. The 1mm scans yield the expected AD cortical thinning pattern~\cite{dickerson2009cortical}, with strong atrophy in the temporal lobe (ES$>$1.0) and, to a lesser extent, in parietal and middle frontal areas (ES$\sim$1.0). The average ES across all regions is 0.64. As expected, the thickness estimates based on the FLAIR scans are less able to detect the differences between the two groups. SynthSR loses, on average, half of the ES (0.32 vs 0.64). Most worrying, it cannot detect the effect on the temporal areas (particularly middle temporal). Our method can detect these differences with ES$>$0.6 in all temporal regions. On average, our method recovers one third of the ES lost by SynthSR (0.42 vs 0.32). 

Finally, we studied the effect of aging on cortical thickness using the same GLM as above. Figure~\ref{fig:ADNI}c shows maps of $p$-values computed with Student's $t$ distribution, where we have transformed $p^* = \log_{10} (p)$ for easier visualization. Once more, the 1mm MPRAGEs display the expected pattern~\cite{salat2004thinning}, with strongest atrophy in superior-temporal and, to less extent, the central and medial frontal gyri. SynthSR fails to detect the superior-temporal effect in the left hemisphere and barely discerns it in the right hemisphere. Our approach, on the other hand, successfully detects these effects. We also note that SynthSR and our method display false positives in frontal areas of the right hemisphere; further analysis (possibly with manual quality control) will be needed to elucidate this result.

 \begin{figure}[t]
\centering
\includegraphics[width=0.88\textwidth]{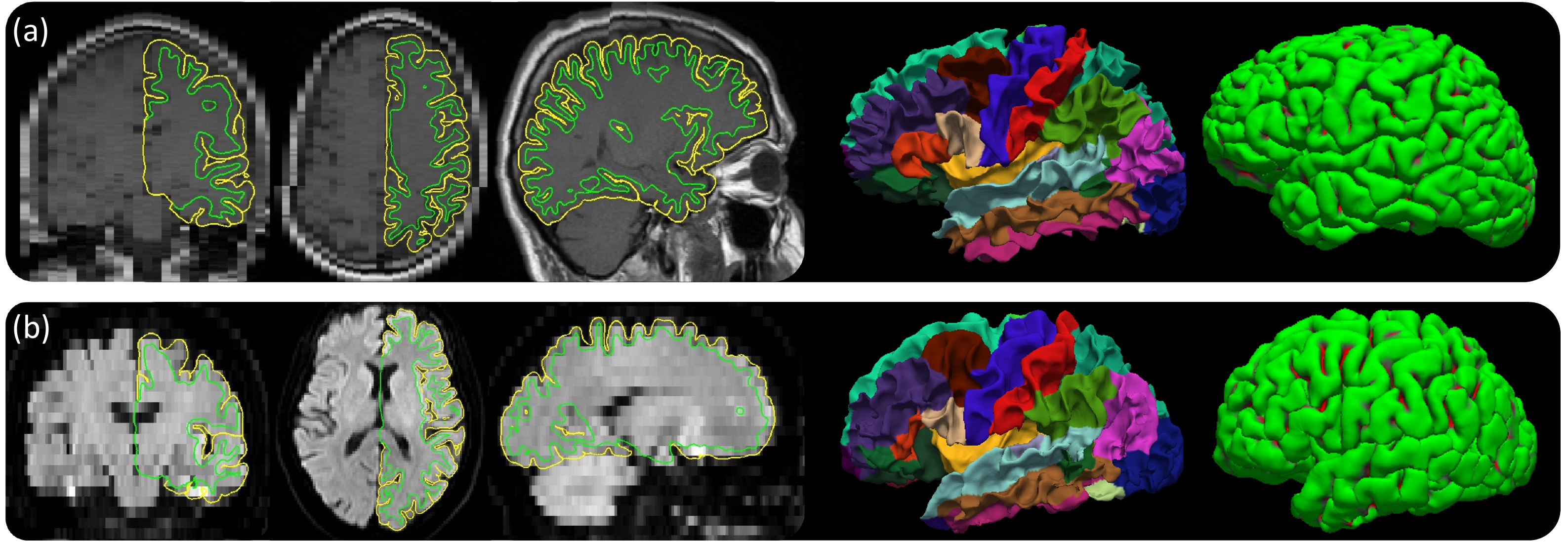}
%\vspace{-5pt}
%https://docs.google.com/drawings/d/1j4vaZMls225DOBuN4bnDE0Ted_jSMhQeyPrfsgUcNRw/edit?usp=sharing
\caption{
Sample outputs for heterogeneous scans from the clinical dataset: (a)~Sagittal TSE-T1 scan (.4$\times$.4$\times$6mm). (b)~Axial FLAIR   (1.7$\times$1.7$\times$6mm). The WM and pial surfaces are shown on the right. The cortical parcellation is overlaid on the WM surface.
}\label{fig:clinical_qualitative}
% \caption{
% Sample outputs for heterogeneous scans from the clinical dataset: (a)~Sagittal TSE-T1 scan with .4$\times$.4$\times$6mm resolution. (b)~Axial FLAIR scan with 1.7$\times$1.7$\times$6mm resolution. (c)~Axial T2-weighted scan with .9$\times$.9$\times$6mm resolution. The left WM and pial surfaces are overlaid on the orthogonal views. The 3D rendered WM and pial surfaces are shown on the right. The cortical parcellation is overlaid on the WM surface.
% }\label{fig:clinical_qualitative}
%\vspace{-5pt}
\end{figure}

\begin{figure}[t]
\centering
\includegraphics[width=0.89\textwidth]{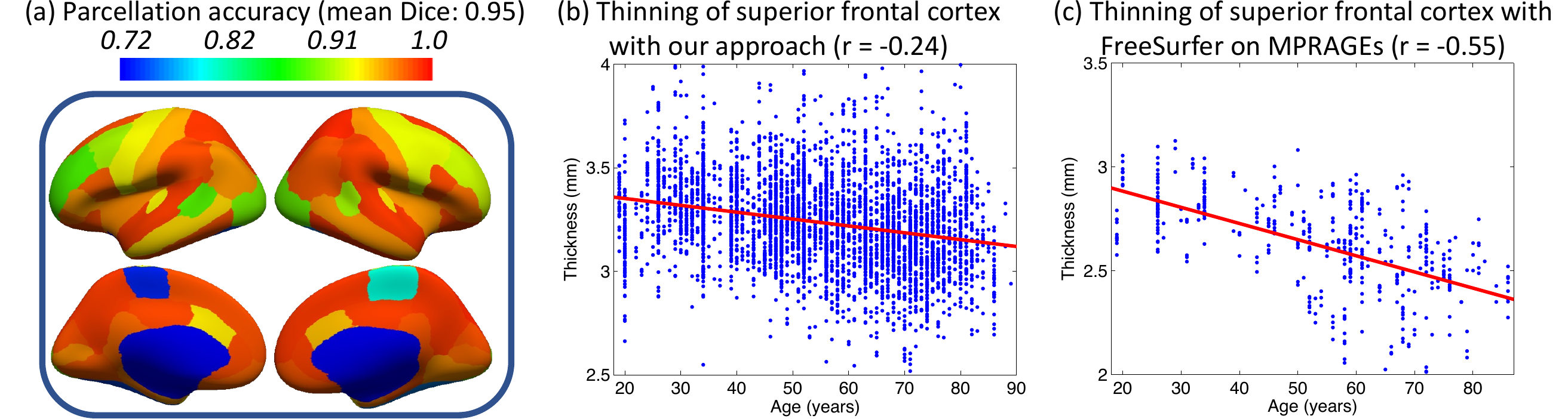}
%\vspace{-5pt}
\caption{
Results on clinical dataset. (a)~Dice scores for parcellation, using FreeSurfer on MPRAGEs as ground truth. (b-c)~Thinning of superior frontal cortex in aging, as measured with our method on anisotropic scans (b) and FreeSurfer on MPRAGEs (c). 
}\label{fig:clinical_quantitative}
%\vspace{-5pt}
\end{figure} 

\subsection{Results on the clinical dataset}

The clinical dataset, despite not being clustered into well defined groups as ADNI, enables us to evaluate our method with the type of data that it is conceived for: a heterogeneous set of brain MRI scans acquired ``in the wild''. Samples of such scans and outputs produced by our method are shown in Figure~\ref{fig:clinical_qualitative}.
In this experiment, we first used the 581 1mm MPRAGEs to compute the Dice scores of the Desikan-Killiany parcellation on clinical acquisitions. The results are displayed in Figure~\ref{fig:clinical_quantitative}a, and show that our proposed method is able to sustain high accuracy in this task (the mean Dice is the same as for ADNI), despite the huge variability in the acquisition protocol of the input scans. As in the previous experiment, we also computed aging curves using all non-1mm-MPRAGE scans (4,483 in total), while correcting for gender and slice spacing. The fitted curve for a representative region (the superior frontal area, which shows consistent effects in Figure~\ref{fig:ADNI}c) is shown in Figure~\ref{fig:clinical_quantitative}b. While the thinning trend exists, the data are rather noisy and the linear fit ($\rho$=-0.24) underestimates the effect of aging, i.e., the magnitude of the slope. This is apparent when comparing with the fit produced by the 581 MPRAGEs  (Figure~\ref{fig:clinical_quantitative}c, $\rho$=-0.55).

%%%%%%%%%%%%%%%%%%%%%%%%%%%%%%%%%%%%%%%%%%%%%%%%%%%%%%%%%%%%%%%%%%%%%%

\subsection{Discussion and conclusion}
We have presented a novel method for cortical analysis of clinical brain scans of any MRI contrast and resolution that does not require retraining. To the best of our knowledge, this is the first method seeking to solve this difficult problem. The method runs in 2-3 hours but could be sped up by replacing some modules (e.g., the spherical registration) with faster learning methods. 

Our method provides accurate parcellation across the board, which is helpful in applications like diffusion MRI (e.g., for seeding or constraining tractography with surfaces and parcellations when a T1 scan is unavailable or is difficult to register due to geometric distortion of the diffusion-weighted images). However,  we observed increased variability in cortical thickness when processing the highly heterogeneous clinical dataset.
Future work will focus on improving the reliability of thickness measurements in such scenarios and assessing if modeling geometric covariates (e.g., vertex-wise distance to the nearest slice or angle between surface and acquisition orientation) may help reduce such variability.

Our method and the clinical dataset are publicly available, which enables researchers worldwide to capitalize on millions of retrospective clinical scans to perform cortical analysis currently unattainable in research studies, particularly for rare diseases and underrepresented populations.

%%%%%%%%%%%%%%%%%%%%%%%%%%%%%%%%%%%%%%%%%%%%%%%%%%%%%%%%%%%%%%%%%%%%%%

\section*{Acknowledgment} This work is primarily funded by the National Institute of Aging (1R01AG070988). Further support is provided by, BRAIN Initiative (1RF1MH123195, 1UM1MH130981), National Institute of Biomedical Imaging and Bioengineering (1R01EB031114), Alzheimer’s Research UK (ARUK-IRG2019A-003), National Institute of Aging (P30AG062421)

% %
% ---- Bibliography ----
%

\bibliographystyle{splncs04}
\bibliography{Reference}

\end{document}